\documentclass[12pt]{article}
\usepackage{amsmath}
\usepackage{amsfonts}
\usepackage{amssymb}
\usepackage{myart}

\normalbaselines
\input tcilatex
\begin{document}

\title{Discrimination of particle masses in multivariant space-time geometry}
\author{Yuri A.Rylov}
\date{Institute for Problems in Mechanics, Russian Academy of Sciences,\\
101-1, Vernadskii Ave., Moscow, 119526, Russia.\\
e-mail: rylov@ipmnet.ru\\
Web site: {$http://rsfq1.physics.sunysb.edu/\symbol{126}rylov/yrylov.htm$}\\
or mirror Web site: {$http://gasdyn-ipm.ipmnet.ru/\symbol{126}%
rylov/yrylov.htm$}}
\maketitle

\begin{abstract}
Multivariance of geometry means that at the point $P_{0}$ there exist many
vectors $\mathbf{P}_{0}\mathbf{P}_{1}$, $\mathbf{P}_{0}\mathbf{P}_{2}$,...
which are equivalent (equal) to the vector $\mathbf{Q}_{0}\mathbf{Q}_{1}$ at
the point $Q_{0}$, but they are not equivalent between themselves. The
discrimination capacity (zero-variance) of geometry appears, when at the
point $P_{0}$ there are no vectors, which are equivalent to the vector $%
\mathbf{Q}_{0}\mathbf{Q}_{1}$ at the point $Q_{0}$. It is shown, that in
some multivariant space-time geometries some particles of small mass may be
discriminated (i.e. either they do not exist, or their evolution is
impossible) . The possibility of some particle discrimination may appear to
be important for explanation of the discrete character of mass spectrum of
elementary particles.
\end{abstract}

\section{Introduction}

Geometrical dynamics is dynamics of elementary particles, generated by the
space-time geometry. In the space-time of Minkowski the geometrical dynamics
coincides with the conventional classical dynamics, and the geometrical
dynamics may be considered to be a generalization of classical dynamics onto
more general space-time geometries. However, the geometric dynamics has a
more fundamental basis, and it may be defined in multivariant space-time
geometries, where one cannot introduce the conventional classical dynamics.
The fact is that, the classical dynamics has been introduced for the
space-time geometry with unlimited divisibility, whereas the real space-time
has a limited divisibility. The limited divisibility of the space-time is of
no importance for dynamics of macroscopic bodies. However, when the size of
moving bodies is of the order of the size of heterogeneity, one may not
neglect the limited divisibility of the space-time geometry.

The geometric dynamics is developed in the framework of the program of the
further physics geometrization, declared in \cite{R2007a}. The special
relativity and the general relativity are steps in the development of this
program. Necessity on the further development appeared in the thirtieth of
the twentieth century, when diffraction of electrons has been discovered.
The motion of electrons, passing through the slit, is multivariant. As far
as the free electron motion depends only on the properties of the
space-time, one needed to change the space-time geometry, making it to be
multivariant. In multivariant geometry at the point $Q_{0}$ there are many
vectors $\mathbf{Q}_{0}\mathbf{Q}_{1}$, $\mathbf{Q}_{0}\mathbf{Q}%
_{1}^{\prime }$,..., which are equal to the given vector $\mathbf{P}_{0}%
\mathbf{P}_{1}$ at the point $P_{0}$, but they are not equal between
themselves. Such a space-time geometry was not known in the beginning of the
twentieth century. It is impossible in the framework of the Riemannian
geometry. As a result the multivariance was prescribed to dynamics. Dynamic
variables were replaced by matrices and operators. One obtains the quantum
dynamics, which differs from the classical dynamics in its principles.
Multivariant space-time geometry appeared only in the end of the twentieth
century \cite{R90,R2002}. The further geometrization of physics became to be
possible.

Any geometry is constructed as a modification of the proper Euclidean
geometry. But not all representations of the proper Euclidean geometry are
convenient for modification. There are three representation of the proper
Euclidean geometry \cite{R2007b}. They differ in the number of primary
(basic) elements, forming the Euclidean geometry.

The Euclidean representation (E-representation) contains three basic
elements (point, segment, angle). Any geometrical object (figure) can be
constructed of these basic elements. Properties of the basic elements and
the method of their application are described by the Euclidean axioms.

The vector representation (V-representation) of the proper Euclidean
geometry contains two basic elements (point, vector). The angle is a
derivative element, which is constructed of two vectors. A use of the two
basic elements at the construction of geometrical objects is determined by
the special structure, known as the linear vector space with the scalar
product, given on it (Euclidean space). The scalar product of linear vector
space describes interrelation of two basic elements (vectors), whereas other
properties of the linear vector space associate with the displacement of
vectors.

The third representation ($\sigma $-representation) of the proper Euclidean
geometry contains only one basic element (point). Segment (vector) is a
derivative element. It is constructed of points. The angle is also a
derivative element. It is constructed of two segments (vectors). The $\sigma 
$-representation contains a special structure: world function $\sigma $,
which describes interrelation of two basic elements (points). The world
function $\sigma \left( P_{0},P_{1}\right) =\frac{1}{2}\rho ^{2}\left(
P_{0},P_{1}\right) $, where $\rho \left( P_{0},P_{1}\right) $ is the
distance between points $P_{0}$ and $P_{1}$. The concept of distance $\rho $%
, as well as the world function $\sigma $, is used in all representations of
the proper Euclidean geometry. However, the world function forms a structure
only in the $\sigma $-representation, where the world function $\sigma $
describes interrelation of two basic elements (points). Besides, the world
function of the proper Euclidean geometry satisfies a series of constraints,
formulated in terms of $\sigma $ and only in terms of $\sigma $. These
conditions (the Euclideaness conditions) will be formulated below.

The Euclideaness conditions are equivalent to a use of the vector linear
space with the scalar product on it, but formally they do not mention the
linear vector space, because all concepts of the linear vector space, as
well as all concepts of the proper Euclidean geometry are expressed directly
via world function $\sigma $ and only via it.

If we want to modify the proper Euclidean geometry, then we should use the $%
\sigma $-representation for its modification. In the $\sigma $%
-representation the special geometric structure (world function) has the
form of a function of two points. Modifying the form of the world function,
we automatically modify all concepts of the proper Euclidean geometry, which
are expressed via the world function. It is very important, that the
expression of geometrical concepts via the world function does not refer to
the means of description (dimension, coordinate system, concept of a curve).
The fact, that modifying the world function, one violates the Euclideaness
conditions, is of no importance, because one obtains non-Euclidean geometry
as a result of such a modification. A change of the world function means a
change of the distance, which is interpreted as a deformation of the proper
Euclidean geometry. The generalized geometry, obtained by a deformation of
the proper Euclidean geometry is called the tubular geometry (T-geometry),
because in the generalized geometry straight lines are tubes (surfaces), in
general, but not one-dimensional lines. Another name of T-geometry is the
physical geometry. The physical geometry is the geometry, described
completely by the world function. Any physical geometry may be used as a
space-time geometry in the sense, that the set of all T-geometries is the
set of all possible space-time geometries.

Modification of the proper Euclidean geometry in V-representation is very
restricted, because in this representation there are two basic elements.
They are not independent, and one cannot modify them independently. Formally
it means, that the linear vector space is to be preserved as a geometrical
structure. It means, in particular, that the generalized geometry retains to
be continuous, uniform and isotropic. The dimension of the generalized
geometry is to be fixed. Besides, the generalized geometry cannot be
multivariant. Such a property of the space-time geometry as multivariance
can be obtained only in $\sigma $-representation. As far as the $\sigma $%
-representation of the proper Euclidean geometry was not known in the
twentieth century, the multivariance of geometry was also unknown concept.

Transition from the V-representation to $\sigma $-representation is carried
out as follows. All concepts of the linear vector space are expressed in
terms of the world function $\sigma $. In reality, concepts of vector,
scalar product of two vectors and linear dependence of $n$ vectors are
expressed via the world function $\sigma _{\mathrm{E}}$ of the proper
Euclidean geometry. Such operations under vectors as equality of vectors,
summation of vectors and multiplication of a vector by a real number are
expressed by means of some formulae. The characteristic properties of these
operations, which are given in V-representation by means of axioms, are
given now by special properties of the Euclidean world function $\sigma _{%
\mathrm{E}}$. After expression of the linear vector space via the world
function the linear vector space may be not mentioned, because all its
properties are described by the world function. We obtain the $\sigma $%
-representation of the proper Euclidean geometry, where some properties of
the linear vector space are expressed in the form of formulae, whereas
another part of properties is hidden in the specific form of the Euclidean
world function $\sigma _{\mathrm{E}}$. Modifying world function, we modify
automatically the properties of the linear vector space (which is not
mentioned in fact). At such a modification we are not to think about the way
of modification of the linear vector space, which is the principal
geometrical structure in the V-representation. In the $\sigma $%
-representation the linear vector space is a derivative structure, which may
be not mention at all. Thus, at transition to $\sigma $-representation the
concepts of the linear vector space (primary concepts in V-representation)
become to be secondary concepts (derivative concepts of the $\sigma $%
-representation).

In $\sigma $-representation we have the following expressions for concepts
of the proper Euclidean geometry. Vector $\mathbf{PQ}=\overrightarrow{PQ}$
is an ordered set of two points $P$ and $Q$. The length $\left\vert \mathbf{%
PQ}\right\vert $ of the vector $\mathbf{PQ}$ is defined by the relation%
\begin{equation}
\left\vert \mathbf{P}_{0}\mathbf{P}_{1}\right\vert =\sqrt{2\sigma \left(
P_{0},P_{1}\right) }  \label{a1.0}
\end{equation}%
The scalar product $\left( \mathbf{P}_{0}\mathbf{P}_{1}.\mathbf{Q}_{0}%
\mathbf{Q}_{1}\right) $ of two vectors $\mathbf{P}_{0}\mathbf{P}_{1}$ and $%
\mathbf{Q}_{0}\mathbf{Q}_{1}$ is defined by the relation%
\begin{equation}
\left( \mathbf{P}_{0}\mathbf{P}_{1}.\mathbf{Q}_{0}\mathbf{Q}_{1}\right)
=\sigma \left( P_{0},Q_{1}\right) +\sigma \left( P_{1},Q_{0}\right) -\sigma
\left( P_{0},Q_{0}\right) -\sigma \left( P_{1},Q_{1}\right)  \label{a1.1}
\end{equation}%
where the world function $\sigma $%
\begin{equation}
\sigma :\qquad \Omega \times \Omega \rightarrow \mathbb{R},\qquad \sigma
\left( P,Q\right) =\sigma \left( Q,P\right) ,\qquad \sigma \left( P,P\right)
=0,\qquad \forall P,Q\in \Omega  \label{a1.2}
\end{equation}%
is the world function $\sigma _{\mathrm{E}}$ of the Euclidean geometry.

In the proper Euclidean geometry $n$ vectors $\mathbf{P}_{0}\mathbf{P}_{k}$, 
$k=1,2,...n$ are linear dependent, if and only if the Gram's determinant 
\begin{equation}
F_{n}\left( \mathcal{P}^{n}\right) =0,\qquad \mathcal{P}^{n}=\left\{
P_{0},P_{1},...,P_{n}\right\}  \label{a1.2a}
\end{equation}%
where the Gram's determinant $F\left( \mathcal{P}^{n}\right) $ is defined by
the relation 
\begin{equation}
F_{n}\left( \mathcal{P}^{n}\right) \equiv \det \left\vert |\left( \mathbf{P}%
_{0}\mathbf{P}_{i}.\mathbf{P}_{0}\mathbf{P}_{k}\right) |\right\vert ,\qquad
i,k=1,2,...n  \label{a1.7}
\end{equation}%
Using expression (\ref{a1.1}) for the scalar product, the condition of the
linear dependence of $n$ vectors $\mathbf{P}_{0}\mathbf{P}_{k}$, $k=1,2,...n$
is written in the form%
\begin{equation}
F_{n}\left( \mathcal{P}^{n}\right) \equiv \det \left\vert |\sigma \left(
P_{0},P_{i}\right) +\sigma \left( P_{0},P_{k}\right) -\sigma \left(
P_{i},P_{k}\right) |\right\vert =0,\qquad i,k=1,2,...n  \label{a1.7b}
\end{equation}

Definition (\ref{a1.1}) of the scalar product of two vectors coincides with
the conventional scalar product of vectors in the proper Euclidean space.
(One can verify this easily). The relations (\ref{a1.1}), (\ref{a1.7b}) do
not contain a reference to the dimension of the Euclidean space and to a
coordinate system in it. Hence, the relations (\ref{a1.1}), (\ref{a1.7b})
are general geometric relations, which may be considered as a definition of
the scalar product of two vectors and that of the linear dependence of
vectors.

Equivalence (equality) of two vectors $\mathbf{P}_{0}\mathbf{P}_{1}$ and $%
\mathbf{Q}_{0}\mathbf{Q}_{1}$ is defined by the relations%
\begin{equation}
\mathbf{P}_{0}\mathbf{P}_{1}\text{eqv}\mathbf{Q}_{0}\mathbf{Q}_{1}:\qquad
\left( \mathbf{P}_{0}\mathbf{P}_{1}.\mathbf{Q}_{0}\mathbf{Q}_{1}\right)
=\left\vert \mathbf{P}_{0}\mathbf{P}_{1}\right\vert \cdot \left\vert \mathbf{%
Q}_{0}\mathbf{Q}_{1}\right\vert \wedge \left\vert \mathbf{P}_{0}\mathbf{P}%
_{1}\right\vert =\left\vert \mathbf{Q}_{0}\mathbf{Q}_{1}\right\vert
\label{a1.3}
\end{equation}%
where $\left\vert \mathbf{P}_{0}\mathbf{P}_{1}\right\vert $ is the length (%
\ref{a1.0}) of the vector $\mathbf{P}_{0}\mathbf{P}_{1}$%
\begin{equation}
\left\vert \mathbf{P}_{0}\mathbf{P}_{1}\right\vert =\sqrt{\left( \mathbf{P}%
_{0}\mathbf{P}_{1}.\mathbf{P}_{0}\mathbf{P}_{1}\right) }=\sqrt{2\sigma
\left( P_{0},P_{1}\right) }  \label{a1.4}
\end{equation}

In the developed form the condition (\ref{a1.3}) of equivalence of two
vectors $\mathbf{P}_{0}\mathbf{P}_{1}$ and $\mathbf{Q}_{0}\mathbf{Q}_{1}$
has the form%
\begin{eqnarray}
\sigma \left( P_{0},Q_{1}\right) +\sigma \left( P_{1},Q_{0}\right) -\sigma
\left( P_{0},Q_{0}\right) -\sigma \left( P_{1},Q_{1}\right) &=&2\sigma
\left( P_{0},P_{1}\right)  \label{a1.5} \\
\sigma \left( P_{0},P_{1}\right) &=&\sigma \left( Q_{0},Q_{1}\right)
\label{a1.6}
\end{eqnarray}

If the points $P_{0},P_{1}$, determining the vector $\mathbf{P}_{0}\mathbf{P}%
_{1}$, and the origin $Q_{0}$ of the vector $\mathbf{Q}_{0}\mathbf{Q}_{1}$
are given, we can determine the vector $\mathbf{Q}_{0}\mathbf{Q}_{1}$, which
is equivalent (equal) to the vector $\mathbf{P}_{0}\mathbf{P}_{1}$, solving
two equations (\ref{a1.5}), (\ref{a1.6}) with respect to the position of the
point $Q_{1}$.

In the case of the proper Euclidean space there is one and only one solution
of equations (\ref{a1.5}), (\ref{a1.6}) independently of the space dimension 
$n$. In the case of arbitrary T-geometry one can guarantee neither existence
nor uniqueness of the solution of equations (\ref{a1.5}), (\ref{a1.6}) for
the point $Q_{1}$. Number of solutions depends on the form of the world
function $\sigma $. This fact means a multivariance of the property of two
vectors equivalence in the arbitrary T-geometry. In other words, the
single-variance of the vector equality in the proper Euclidean space is a
specific property of the proper Euclidean geometry, and this property is
conditioned by the form of the Euclidean world function. In other
T-geometries this property does not take place, in general.

The multivariance is a general property of a physical geometry. It is
connected with a necessity of solution of algebraic equations, containing
the world function. As far as the world function is different in different
physical geometries, the solution of these equations may be not unique, or
it may not exist at all. If there are many solutions of equations (\ref{a1.5}%
), (\ref{a1.6}) at fixed vector $\mathbf{P}_{0}\mathbf{P}_{1}$ and fixed
point $Q_{0}$, we shall speak on the property of multivariance of the
physical geometry. If there is no solution of equations (\ref{a1.5}), (\ref%
{a1.6}) at fixed vector $\mathbf{P}_{0}\mathbf{P}_{1}$ and fixed point $%
Q_{0} $, we shall speak, that the physical geometry has the property of
discrimination (zero-variance).

If in the $n$-dimensional Euclidean space $F_{n}\left( \mathcal{P}%
^{n}\right) \neq 0$, the vectors $\mathbf{P}_{0}\mathbf{P}_{k}$, $k=1,2,...n$
are linear independent. We may construct rectilinear coordinate system with
basic vectors $\mathbf{P}_{0}\mathbf{P}_{k}$, $k=1,2,...n$ in the $n$%
-dimensional Euclidean space. Covariant coordinates $x_{k}=\left( \mathbf{P}%
_{0}\mathbf{P}\right) _{k}$ of the vector $\mathbf{P}_{0}\mathbf{P}$ in this
coordinate system have the form%
\begin{equation}
x_{k}=x_{k}\left( P\right) =\left( \mathbf{P}_{0}\mathbf{P}\right)
_{k}=\left( \mathbf{P}_{0}\mathbf{P}.\mathbf{P}_{0}\mathbf{P}_{k}\right)
,\qquad k=1,2,...n  \label{a1.8}
\end{equation}

Now we can formulate the Euclideaness conditions. These conditions are
conditions of the fact, that the T-geometry, described by the world function 
$\sigma $, is $n$-dimensional proper Euclidean geometry.

I. Definition of the dimension and introduction of the rectilinear
coordinate system: 
\begin{equation}
\exists \mathcal{P}^{n}\equiv \left\{ P_{0},P_{1},...P_{n}\right\} \subset
\Omega ,\qquad F_{n}\left( \mathcal{P}^{n}\right) \neq 0,\qquad F_{k}\left( {%
\Omega }^{k+1}\right) =0,\qquad k>n  \label{g2.5}
\end{equation}%
where $F_{n}\left( \mathcal{P}^{n}\right) $\ is the Gram's determinant (\ref%
{a1.7}). Vectors $\mathbf{P}_{0}\mathbf{P}_{i}$, $\;i=1,2,...n$\ are basic
vectors of the rectilinear coordinate system $K_{n}$\ with the origin at the
point $P_{0}$. In $K_{n}$ the covariant metric tensor $g_{ik}\left( \mathcal{%
P}^{n}\right) $, \ $i,k=1,2,...n$\ and the contravariant one $g^{ik}\left( 
\mathcal{P}^{n}\right) $, \ $i,k=1,2,...n$\ \ are defined by the relations 
\begin{equation}
\sum\limits_{k=1}^{k=n}g^{ik}\left( \mathcal{P}^{n}\right) g_{lk}\left( 
\mathcal{P}^{n}\right) =\delta _{l}^{i},\qquad g_{il}\left( \mathcal{P}%
^{n}\right) =\left( \mathbf{P}_{0}\mathbf{P}_{i}.\mathbf{P}_{0}\mathbf{P}%
_{l}\right) ,\qquad i,l=1,2,...n  \label{a1.5b}
\end{equation}%
\begin{equation}
F_{n}\left( \mathcal{P}^{n}\right) =\det \left\vert \left\vert g_{ik}\left( 
\mathcal{P}^{n}\right) \right\vert \right\vert \neq 0,\qquad i,k=1,2,...n
\label{g2.6}
\end{equation}

II. Linear structure of the Euclidean space: 
\begin{equation}
\sigma \left( P,Q\right) =\frac{1}{2}\sum\limits_{i,k=1}^{i,k=n}g^{ik}\left( 
\mathcal{P}^{n}\right) \left( x_{i}\left( P\right) -x_{i}\left( Q\right)
\right) \left( x_{k}\left( P\right) -x_{k}\left( Q\right) \right) ,\qquad
\forall P,Q\in \Omega  \label{a1.5a}
\end{equation}%
where coordinates $x_{i}=x_{i}\left( P\right) ,$\ $i=1,2,...n$\ of the point 
$P$\ are covariant coordinates of the vector $\mathbf{P}_{0}\mathbf{P}$,
defined by the relation (\ref{a1.8}).

III: The metric tensor matrix $g_{lk}\left( \mathcal{P}^{n}\right) $\ has
only positive eigenvalues 
\begin{equation}
g_{k}>0,\qquad k=1,2,...,n  \label{a15c}
\end{equation}

IV. The continuity condition: the system of equations 
\begin{equation}
\left( \mathbf{P}_{0}\mathbf{P}_{i}.\mathbf{P}_{0}\mathbf{P}\right)
=y_{i}\in \mathbb{R},\qquad i=1,2,...n  \label{b14}
\end{equation}%
considered to be equations for determination of the point $P$\ as a function
of coordinates $y=\left\{ y_{i}\right\} $,\ \ $i=1,2,...n$\ has always one
and only one solution.\textit{\ }All conditions I $\div $ IV contain a
reference to the dimension $n$\ of the Euclidean space.

One can show that conditions I $\div $ IV are the necessary and sufficient
conditions of the fact, that the world function $\sigma $, given on $\Omega $%
, describes the $n$-dimensional Euclidean space \cite{R90}.

\section{Dynamics as a result of the space-time \newline
geometry}

Construction of dynamics in the space-time on the basis of a physical
geometry (T-geometry), is presented in \cite{R2007a}. Here we remind the
statement of the problem of dynamics.

Geometrical object $\mathcal{O\subset }\Omega $ is a subset of points in the
point set $\Omega $. In the T-geometry the geometric object $\mathcal{O}$ is
described by means of the skeleton-envelope method. It means that any
geometric object $\mathcal{O}$ is considered to be a set of intersections
and joins of elementary geometric objects (EGO).

The elementary geometrical object $\mathcal{E}$ is described by its skeleton 
$\mathcal{P}^{n}$ and envelope function $f_{\mathcal{P}^{n}}$. The finite
set $\mathcal{P}^{n}\equiv \left\{ P_{0},P_{1},...,P_{n}\right\} \subset
\Omega $ of parameters of the envelope function $f_{\mathcal{P}^{n}}$ is the
skeleton of elementary geometric object (EGO) $\mathcal{E}\subset \Omega $.
The set $\mathcal{E}\subset \Omega $ of points forming EGO is called the
envelope of its skeleton $\mathcal{P}^{n}$. The envelope function $f_{%
\mathcal{P}^{n}}$%
\begin{equation}
f_{\mathcal{P}^{n}}:\qquad \Omega \rightarrow \mathbb{R},  \label{h2.1}
\end{equation}%
determining EGO is a function of the running point $R\in \Omega $ and of
parameters $\mathcal{P}^{n}\subset \Omega $. The envelope function $f_{%
\mathcal{P}^{n}}$ is supposed to be an algebraic function of $s$ arguments $%
w=\left\{ w_{1},w_{2},...w_{s}\right\} $, $s=(n+2)(n+1)/2$. Each of
arguments $w_{k}=\sigma \left( Q_{k},L_{k}\right) $ is the world function $%
\sigma $ of two points $Q_{k},L_{k}\in \left\{ R,\mathcal{P}^{n}\right\} $,
either belonging to skeleton $\mathcal{P}^{n}$, or coinciding with the
running point $R$. Thus, any elementary geometric object $\mathcal{E}$ is
determined by its skeleton $\mathcal{P}^{n}$ and its envelope function $f_{%
\mathcal{P}^{n}}$. Elementary geometric object $\mathcal{E}$ is the set of
zeros of the envelope function 
\begin{equation}
\mathcal{E}=\left\{ R|f_{\mathcal{P}^{n}}\left( R\right) =0\right\}
\label{h2.2}
\end{equation}%
\textit{Definition.} Two EGOs $\mathcal{E}_{\mathcal{P}^{n}}$ and $\mathcal{E%
}_{\mathcal{Q}^{n}}$ are equivalent, if their skeletons $\mathcal{P}^{n}$
and $\mathcal{Q}^{n}$ are equivalent and their envelope functions $f_{%
\mathcal{P}^{n}}$ and $g_{\mathcal{Q}^{n}}$ are equivalent. Equivalence ($%
\mathcal{P}^{n}$eqv$\mathcal{Q}^{n}$) of two skeletons $\mathcal{P}%
^{n}\equiv \left\{ P_{0},P_{1},...,P_{n}\right\} \subset \Omega $ and $%
\mathcal{Q}^{n}\equiv \left\{ Q_{0},Q_{1},...,Q_{n}\right\} \subset \Omega $
means that 
\begin{equation}
\mathcal{P}^{n}\text{eqv}\mathcal{Q}^{n}:\qquad \mathbf{P}_{i}\mathbf{P}_{k}%
\text{eqv}\mathbf{Q}_{i}\mathbf{Q}_{k},\qquad i,k=0,1,...n,\quad i\leq k
\label{a5.4}
\end{equation}%
Equivalence of the envelope functions $f_{\mathcal{P}^{n}}$ and $g_{\mathcal{%
Q}^{n}}$ means, that they have the same set of zeros. It means that 
\begin{equation}
f_{\mathcal{P}^{n}}\left( R\right) =\Phi \left( g_{\mathcal{P}^{n}}\left(
R\right) \right) ,\qquad \forall R\in \Omega  \label{a5.5}
\end{equation}%
where $\Phi $ is an arbitrary function, having the property%
\begin{equation}
\Phi :\mathbb{R}\rightarrow \mathbb{R},\qquad \Phi \left( 0\right) =0
\label{a5.5a}
\end{equation}

Evolution of EGO $\mathcal{O}_{\mathcal{P}^{n}}$ in the space-time is
described as a world chain $\mathcal{C}_{\mathrm{fr}}$ of equivalent
connected EGOs. The point $P_{0}$ of the skeleton $\mathcal{P}^{n}=\left\{
P_{0},P_{1},...P_{n}\right\} $ is considered to be the origin of the
geometrical object $\mathcal{O}_{\mathcal{P}^{n}}.$ The EGO $\mathcal{O}_{%
\mathcal{P}^{n}}$ is considered to be placed at its origin $P_{0}$. Let us
consider a set of equivalent skeletons $\mathcal{P}_{\left( l\right)
}^{n}=\left\{ P_{0}^{\left( l\right) },P_{1}^{\left( l\right)
},...P_{n}^{\left( l\right) }\right\} ,$ $l=...0,1,...$which are equivalent
in pairs 
\begin{equation}
\mathbf{P}_{i}^{\left( l\right) }\mathbf{P}_{k}^{\left( l\right) }\text{eqv}%
\mathbf{P}_{i}^{\left( l+1\right) }\mathbf{P}_{k}^{\left( l+1\right)
},\qquad i,k=0,1,...n;\qquad l=...1,2,...  \label{a6.1}
\end{equation}%
The skeletons $\mathcal{P}_{\left( l\right) }^{n},$ $l=...0,1,...$are
connected, and they form a chain in the direction of vector $\mathbf{P}_{0}%
\mathbf{P}_{1}$, if the point $P_{1}$ of one skeleton coincides with the
origin $P_{0}$ of the adjacent skeleton 
\begin{equation}
P_{1}^{\left( l\right) }=P_{0}^{\left( l+1\right) },\qquad l=...0,1,2,...
\label{a6.2}
\end{equation}%
The chain $\mathcal{C}_{\mathrm{fr}}$ describes evolution of the elementary
geometrical object $\mathcal{O}_{\mathcal{P}^{n}}$ in the direction of the
leading vector $\mathbf{P}_{0}\mathbf{P}_{1}$. The evolution of EGO $%
\mathcal{O}_{\mathcal{P}^{n}}$ is a temporal evolution, if the vectors $%
\mathbf{P}_{0}^{\left( l\right) }\mathbf{P}_{1}^{\left( l\right) }$ are
timelike $\left\vert \mathbf{P}_{0}^{\left( l\right) }\mathbf{P}_{1}^{\left(
l\right) }\right\vert >0,$ $\ l=...0,1,..$. The evolution of EGO $\mathcal{O}%
_{\mathcal{P}^{n}}$ is a spatial evolution, if the vectors $\mathbf{P}%
_{0}^{\left( l\right) }\mathbf{P}_{1}^{\left( l\right) }$ are spacelike $%
\left\vert \mathbf{P}_{0}^{\left( l\right) }\mathbf{P}_{1}^{\left( l\right)
}\right\vert <0,$ $\ l=...0,1,..$.

Note, that all adjacent links (EGOs) of the chain are equivalent in pairs,
although two links of the chain may be not equivalent, if they are not
adjacent. However, lengths of corresponding vectors are equal in all links
of the chain 
\begin{equation}
\left\vert \mathbf{P}_{i}^{\left( l\right) }\mathbf{P}_{k}^{\left( l\right)
}\right\vert =\left\vert \mathbf{P}_{i}^{\left( s\right) }\mathbf{P}%
_{k}^{\left( s\right) }\right\vert ,\qquad i,k=0,1,...n;\qquad l,s=...1,2,...
\label{a6.4}
\end{equation}%
We shall refer to the vector $\mathbf{P}_{0}^{\left( l\right) }\mathbf{P}%
_{1}^{\left( l\right) }$, which determines the form of the evolution and the
shape of the world chain, as the leading vector. This vector determines the
direction of 4-velocity of the physical body, associated with the link of
the world chain.

If the relations%
\begin{eqnarray}
\mathcal{P}^{n}\mathrm{eqv}\mathcal{Q}^{n} &:&\qquad \left( \mathbf{P}_{i}%
\mathbf{P}_{k}.\mathbf{Q}_{i}\mathbf{Q}_{k}\right) =\left\vert \mathbf{P}_{i}%
\mathbf{P}_{k}\right\vert \cdot \left\vert \mathbf{Q}_{i}\mathbf{Q}%
_{k}\right\vert ,\qquad \left\vert \mathbf{P}_{i}\mathbf{P}_{k}\right\vert
=\left\vert \mathbf{Q}_{i}\mathbf{Q}_{k}\right\vert ,  \label{a6.9} \\
i,k &=&0,1,2,...n  \notag
\end{eqnarray}%
\begin{eqnarray}
\mathcal{Q}^{n}\mathrm{eqv}\mathcal{R}^{n} &:&\qquad \left( \mathbf{Q}_{i}%
\mathbf{Q}_{k}.\mathbf{R}_{i}\mathbf{R}_{k}\right) =\left\vert \mathbf{Q}_{i}%
\mathbf{Q}_{k}\right\vert \cdot \left\vert \mathbf{R}_{i}\mathbf{R}%
_{k}\right\vert ,\qquad \left\vert \mathbf{Q}_{i}\mathbf{Q}_{k}\right\vert
=\left\vert \mathbf{R}_{i}\mathbf{R}_{k}\right\vert ,  \label{a6.10} \\
i,k &=&0,1,2,...n  \notag
\end{eqnarray}%
are satisfied, the relations%
\begin{eqnarray}
\mathcal{P}^{n}\mathrm{eqv}\mathcal{R}^{n} &:&\qquad \left( \mathbf{P}_{i}%
\mathbf{P}_{k}.\mathbf{R}_{i}\mathbf{R}_{k}\right) =\left\vert \mathbf{P}_{i}%
\mathbf{P}_{k}\right\vert \cdot \left\vert \mathbf{R}_{i}\mathbf{R}%
_{k}\right\vert ,\qquad \left\vert \mathbf{P}_{i}\mathbf{P}_{k}\right\vert
=\left\vert \mathbf{R}_{i}\mathbf{R}_{k}\right\vert ,  \label{a6.11} \\
i,k &=&0,1,2,...n  \notag
\end{eqnarray}%
are not satisfied, in general, because the relations (\ref{a6.11}) contain
the scalar products $\left( \mathbf{P}_{i}\mathbf{P}_{k}.\mathbf{R}_{i}%
\mathbf{R}_{k}\right) $. These scalar products contain the world functions $%
\sigma \left( P_{i},R_{k}\right) $, which are not contained in relations (%
\ref{a6.9}), (\ref{a6.10}).

The world chain $\mathcal{C}_{\mathrm{fr}}$, consisting of equivalent links (%
\ref{a6.1}), (\ref{a6.2}), describes a free motion of a physical body
(particle), associated with the skeleton $\mathcal{P}^{n}$. We assume that 
\textit{the motion of a physical body is free, if all points of the body
move free} (i.e. without acceleration). If the external forces are absent,
the physical body as a whole moves without acceleration. However, if the
body rotates, one may not consider the motion of this body as a free motion,
because not all points of this body move free (without acceleration). In the
rotating body there are internal forces, which generate centripetal
acceleration to some points of the body. As a result some points of the body
do not move free. Motion of the rotating body may be free only on the
average, but not exactly free.

Conception of non-free motion of a particle is rather indefinite, and we
restrict ourselves only with consideration of a free motion.

Conventional conception of the motion of extensive (non-pointlike) particle,
which is free on the average, contains a free displacement, described by the
velocity 4-vector, and a spatial rotation, described by the angular velocity
3-pseudovector $\mathbf{\Omega }$. The velocity 4-vector is associated with
the timelike leading vector $\mathbf{P}_{0}\mathbf{P}_{1}$. At the free on
the average motion of a rotating body some of vectors $\mathbf{P}_{0}\mathbf{%
P}_{2}^{\mathrm{(s)}},\mathbf{P}_{0}\mathbf{P}_{3}^{\mathrm{(s)}},$...of the
skeleton $\mathcal{P}^{n}$ are not in parallel with vectors $\mathbf{P}_{0}%
\mathbf{P}_{2}^{\mathrm{(s+1)}},\mathbf{P}_{0}\mathbf{P}_{3}^{\mathrm{(s+1)}%
},...$, although at the free motion all vectors $\mathbf{P}_{0}\mathbf{P}%
_{2}^{\mathrm{(s)}},\mathbf{P}_{0}\mathbf{P}_{3}^{\mathrm{(s)}},...$ are to
be in parallel with $\mathbf{P}_{0}\mathbf{P}_{2}^{\mathrm{(s+1)}},\mathbf{P}%
_{0}\mathbf{P}_{3}^{\mathrm{(s+1)}},...$ as it follows from (\ref{a6.1}). It
means that the world chain $\mathcal{C}_{\mathrm{fr}}$ of a freely moving
body can describe only translation of a physical body, but not its rotation.

If the leading vector $\mathbf{P}_{0}\mathbf{P}_{1}$ is spacelike, the body,
described by the skeleton $\mathcal{P}^{n}$, evolves in the spacelike
direction. It seems, that the spacelike evolution is prohibited. But it is
not so. If the world chain forms a helix with the timelike axis, such a
world chain may be considered as timelike on the average. In reality such
world chains are possible. For instance, the world chain of the classical
Dirac particle is a helix with timelike axis \cite{R2001c,R2004,R2004b}. It
is not quite clear, whether or not the links of this chain are spacelike,
because internal degrees of freedom of the Dirac particle, responsible for
helicity of the world chain, are described nonrelativistically.

Thus, consideration of a spatial evolution is not meaningless, especially if
we take into account, that the spatial evolution may imitate rotation, which
is absent at the free motion of a particle.

\section{Discreteness and zero-variance of multivariant space-time geometry}

Let us consider the flat homogeneous isotropic space-time $V_{\mathrm{d}%
}=\left\{ \sigma _{\mathrm{d}},\mathbb{R}^{4}\right\} $, described by the
world function 
\begin{equation}
\sigma _{\mathrm{d}}=\sigma _{\mathrm{M}}+d\cdot \mathrm{sgn}\left( \sigma _{%
\mathrm{M}}\right) ,\qquad d=\lambda _{0}^{2}=\text{const}>0  \label{a4.0}
\end{equation}%
\begin{equation}
\mathrm{sgn}\left( x\right) =\left\{ 
\begin{array}{l}
1,\ \ \text{if}\ \ x>0 \\ 
0,\ \ \ \ \text{if\ \ }x=0 \\ 
-1,\ \ \text{if}\ \ x<0%
\end{array}%
\right. ,  \label{a4.1}
\end{equation}%
where $\sigma _{\mathrm{M}}$ is the world function of the $4$-dimensional
space-time of Minkowski. $\lambda _{0}$ is some elementary length. In such a
space-time geometry two connected equivalent timelike vectors $\mathbf{P}_{0}%
\mathbf{P}_{1}$ and $\mathbf{P}_{1}\mathbf{P}_{2}$ are described as follows 
\cite{R2007a} 
\begin{equation}
\mathbf{P}_{0}\mathbf{P}_{1}\mathrm{eqv}\mathbf{P}_{1}\mathbf{P}_{2}:\qquad 
\mathbf{P}_{0}\mathbf{P}_{1}=\left\{ \mu ,0,0,0\right\} ,\qquad \mathbf{P}%
_{1}\mathbf{P}_{2}=\left\{ \mu +\frac{3\lambda _{0}^{2}}{\mu },\lambda _{0}%
\sqrt{6+\frac{9\lambda _{0}^{2}}{\mu ^{2}}}\mathbf{n}\right\}  \label{a6.15}
\end{equation}%
where $\mathbf{n=}\left\{ n_{1},n_{2},n_{3}\right\} $ is an arbitrary unit
3-vector ($n_{1}^{2}+n_{2}^{2}+n_{3}^{2}=1$). The quantity $\mu $ is the
length of the vector $\mathbf{P}_{0}\mathbf{P}_{1}$ (geometrical mass,
associated with the particle, which is described by the vector $\mathbf{P}%
_{0}\mathbf{P}_{1}$). We see that the spatial part of the vector $\mathbf{P}%
_{1}\mathbf{P}_{2}$ is determined to within the arbitrary 3-vector of the
length $\lambda _{0}\sqrt{6+\frac{9\lambda _{0}^{2}}{\mu ^{2}}}$. This
multivariance generates wobbling of the links of the world chain, consisting
of equivalent timelike vectors $...\mathbf{P}_{0}\mathbf{P}_{1}$, $\mathbf{P}%
_{1}\mathbf{P}_{2}$, $\mathbf{P}_{2}\mathbf{P}_{3}$,... Statistical
description of the chain with wobbling links coincides with the quantum
description of the particle with the mass $m=b\mu $, if the elementary
length $\lambda _{0}=\hbar ^{1/2}\left( 2bc\right) ^{-1/2}$, where $c$ is
the speed of the light, $\hbar $ is the quantum constant, and $b$ is some
universal constant, whose exact value is not determined \cite{R91}, because
the statistical description does not contain the quantity $b$. Thus, the
characteristic wobbling length is of the order of $\lambda _{0}$.

Thus, to explain the quantum description of the particle motion as a
statistical description of the multivariant classical motion, we should use
the world function (\ref{a4.0}). However, the form of the world function (%
\ref{a4.0}) is determined by the coincidence of the two description only for
the value $\sigma _{\mathrm{M}}>\sigma _{0}$, where the constant $\sigma
_{0} $ is determined via the mass $m_{\mathrm{L}}$ of the lightest massive
particle (electron) by means of the relation 
\begin{equation}
\sigma _{0}\leq \frac{\mu _{\mathrm{L}}^{2}}{2}-d=\frac{m_{\mathrm{L}}^{2}}{%
2b^{2}}-d=\frac{m_{\mathrm{L}}^{2}}{2b^{2}}-\frac{\hbar }{2bc}  \label{a4.2}
\end{equation}%
where $\mu _{\mathrm{L}}=m_{\mathrm{L}}/b$ is the geometrical mass of the
lightest massive particle (electron). The geometrical mass $\mu _{\mathrm{LM}%
}$ of the same particle, considered in the space-time geometry of Minkowski,
has the form%
\begin{equation*}
\mu _{\mathrm{LM}}=\sqrt{\mu _{\mathrm{L}}^{2}-2d}
\end{equation*}%
As far as $\sigma _{0}>0$, and, hence, $m_{\mathrm{L}}^{2}-b\hbar c^{-1}>0$,
we obtain the following estimation for the universal constant $b$%
\begin{equation}
b<\frac{m_{\mathrm{L}}^{2}c}{\hbar }\approx 2.4\times 10^{-17}\text{g/cm}.
\label{a4.3}
\end{equation}

Intensity of wobbling may be described by the multivariance vector $\mathbf{a%
}_{\mathrm{m}}$, which is defined as follows. Let $\mathbf{P}_{1}\mathbf{P}%
_{2}$, $\mathbf{P}_{1}\mathbf{P}_{2}^{\prime }$ be to vectors which are
equivalent to the vector $\mathbf{P}_{0}\mathbf{P}_{1}$. Let 
\begin{equation*}
\mathbf{P}_{1}\mathbf{P}_{2}=\left\{ \mu +\frac{3\lambda _{0}^{2}}{\mu }%
,\lambda _{0}\sqrt{6+\frac{9\lambda _{0}^{2}}{\mu ^{2}}}\mathbf{n}\right\}
,\qquad \mathbf{P}_{1}\mathbf{P}_{2}^{\prime }=\left\{ \mu +\frac{3\lambda
_{0}^{2}}{\mu },\lambda _{0}\sqrt{6+\frac{9\lambda _{0}^{2}}{\mu ^{2}}}%
\mathbf{n}^{\prime }\right\}
\end{equation*}%
Let us consider the vector 
\begin{equation}
\mathbf{P}_{2}\mathbf{P}_{2}^{\prime }=\left\{ 0,\lambda _{0}\sqrt{6+\frac{%
9\lambda _{0}^{2}}{\mu ^{2}}}\left( \mathbf{n}^{\prime }-\mathbf{n}\right)
\right\}  \label{a4.3a}
\end{equation}%
which is a difference of vectors $\mathbf{P}_{1}\mathbf{P}_{2}$, $\mathbf{P}%
_{1}\mathbf{P}_{2}^{\prime }$. We consider the length $\left\vert \mathbf{P}%
_{2}\mathbf{P}_{2}^{\prime }\right\vert _{\mathrm{M}}$ of the vector $%
\mathbf{P}_{2}\mathbf{P}_{2}^{\prime }$ in the Minkowski space-time. We
obtain%
\begin{equation}
\left\vert \mathbf{P}_{2}\mathbf{P}_{2}^{\prime }\right\vert _{\mathrm{M}%
}^{2}=-\lambda _{0}^{2}\left( 6+\frac{9\lambda _{0}^{2}}{\mu ^{2}}\right)
\left( 2-2\mathbf{nn}^{\prime }\right)  \label{a4.3b}
\end{equation}%
The length of the vector (\ref{a4.3a}) is minimal at $\mathbf{n}=-\mathbf{n}%
^{\prime }$. At $\mathbf{n}=\mathbf{n}^{\prime }$ the length of the vector (%
\ref{a4.3a}) is maximal, and it is equal to zero. By definition the vector $%
\mathbf{P}_{2}\mathbf{P}_{2}^{\prime }$ at $\mathbf{n}=-\mathbf{n}^{\prime }$
is the multivariance 4-vector $\mathbf{a}_{\mathrm{m}}$, which describes the
intensity of the multivariance. We have 
\begin{equation}
\mathbf{a}_{\mathrm{m}}=\left\{ 0,\ 2\lambda _{0}\sqrt{6+\frac{9\lambda
_{0}^{2}}{\mu ^{2}}}\mathbf{n}\right\} \qquad \left\vert \mathbf{a}_{\mathrm{%
m}}\right\vert ^{2}=\left( \mathbf{a}_{\mathrm{m}}.\mathbf{a}_{\mathrm{m}%
}\right) =-4\lambda _{0}^{2}\left( 6+\frac{9\lambda _{0}^{2}}{\mu ^{2}}%
\right)  \label{a4.3c}
\end{equation}%
where $\mathbf{n}$ is an arbitrary unit 3-vector. The multivariance vector $%
\mathbf{a}_{\mathrm{m}}$ is spacelike

In the case, when $\mu \gg \lambda _{0}$, corresponding wobbling length 
\begin{equation*}
\lambda _{\mathrm{w}}=\frac{1}{2}\sqrt{\left\vert \left( \mathbf{a}_{\mathrm{%
m}}.\mathbf{a}_{\mathrm{m}}\right) \right\vert }\approx \sqrt{6}\lambda _{0}=%
\sqrt{6}\sqrt{\frac{\hbar }{2bc}}>\sqrt{3}\frac{\hbar }{m_{\mathrm{L}}c}=%
\sqrt{3}\lambda _{\mathrm{com}}
\end{equation*}%
where $\lambda _{\mathrm{com}}$ is the Compton wave length of electron.

The relation (\ref{a4.3}) means that 
\begin{equation}
\sigma _{\mathrm{d}}=\sigma _{\mathrm{M}}+d,\ \ \ \ \text{if }\sigma _{%
\mathrm{M}}>\sigma _{0}  \label{a4.4}
\end{equation}%
For other values $\sigma _{\mathrm{M}}<\sigma _{0}$ the form of the world
function $\sigma _{\mathrm{d}}$ may distinguish from the relation (\ref{a4.4}%
). However, $\sigma _{\mathrm{d}}=0$, if $\sigma _{\mathrm{M}}=0$.

The space-time (\ref{a4.0}) is a discrete space-time, $\left\vert \mathbf{P}%
_{0}\mathbf{P}_{1}\right\vert ^{2}\geq \lambda _{0}^{2}$ , if the vector $%
\mathbf{P}_{0}\mathbf{P}_{1}$ is timelike, and $\left\vert \mathbf{P}_{0}%
\mathbf{P}_{1}\right\vert ^{2}\leq \lambda _{0}^{2}$ , if the vector $%
\mathbf{P}_{0}\mathbf{P}_{1}$ is spacelike. It means that in the space-time
there are no close points. Absence of close points on the continual set of
points seems rather unexpected, because the space-time discreteness
associates usually with a latticed space-time, but not with a continual
space-time. Nevertheless, the fact of absence of close points means a
discreteness of the space-time, and we see no other interpretation, than
discreteness. If we consider equivalence of two vectors $\mathbf{P}_{0}%
\mathbf{P}_{1}$ and $\mathbf{Q}_{0}\mathbf{Q}_{1}$ in some latticed space,
we should expect a discrimination in some cases, because due to latticed
character of the space one cannot always find at the point $P_{0}$ a vector $%
\mathbf{P}_{0}\mathbf{P}_{1}$, which is equivalent to a given vector $%
\mathbf{Q}_{0}\mathbf{Q}_{1}$.

Let us consider the space-time, described by the world function%
\begin{equation}
\sigma =\sigma _{\mathrm{M}}+d\left( \sigma _{\mathrm{M}}\right)
\label{a4.5}
\end{equation}%
\begin{equation}
d\left( \sigma _{\mathrm{M}}\right) =\lambda _{0}^{2}f\left( \frac{\sigma _{%
\mathrm{M}}}{\sigma _{0}}\right) =\left\{ 
\begin{array}{lll}
\lambda _{0}^{2}\mathrm{sgn}\left( \frac{\sigma _{\mathrm{M}}}{\sigma _{0}}%
\right) & \text{if} & \left\vert \sigma _{\mathrm{M}}\right\vert >\sigma
_{0}>0 \\ 
\lambda _{0}^{2}\frac{\sigma _{\mathrm{M}}}{\sigma _{0}} & \text{if} & 
\left\vert \sigma _{\mathrm{M}}\right\vert \leq \sigma _{0}%
\end{array}%
\right.  \label{a4.6}
\end{equation}

If $\sigma _{0}$ is small, the world function is close to the world function
(\ref{a4.0}). If $\sigma _{0}\rightarrow 0$, the world function (\ref{a4.6})
tends to (\ref{a4.0}). Strictly, the space-time geometry (\ref{a4.6}) is not
discrete, however it is close to the discrete space-time geometry (\ref{a4.0}%
). This semidiscreteness manifests itself in a the capacity of
discrimination, when evolution of some particles with small mass appears to
be prohibited.

Let us consider two connected timelike vectors $\mathbf{P}_{0}\mathbf{P}_{1}$%
, $\mathbf{P}_{1}\mathbf{P}_{2}$ of the world chain $\mathbf{P}_{k}\mathbf{P}%
_{k+1}$, $k=...0,1,...$ Let us use the inertial coordinate system, where the
points $P_{0},P_{1},P_{2}$ have coordinates%
\begin{equation}
P_{0}=\left\{ 0,0,0,0\right\} ,\qquad P_{1}=\left\{ s,0,0,0\right\} ,\qquad
P_{2}=\left\{ 2s+\alpha _{0},\alpha _{1},\alpha _{2},\alpha _{3}\right\}
\label{a4.7}
\end{equation}%
and coordinates of corresponding vectors are%
\begin{eqnarray}
\mathbf{P}_{0}\mathbf{P}_{1} &=&\left\{ s,0,0,0\right\} ,\qquad \mathbf{P}%
_{1}\mathbf{P}_{2}=\left\{ s+\alpha _{0},\alpha _{1},\alpha _{2},\alpha
_{3}\right\}  \label{a4.8} \\
\mathbf{P}_{0}\mathbf{P}_{2} &=&\left\{ 2s+\alpha _{0},\alpha _{1},\alpha
_{2},\alpha _{3}\right\}  \label{a4.9}
\end{eqnarray}%
In this case 
\begin{equation}
\left\vert \mathbf{P}_{0}\mathbf{P}_{1}\right\vert _{\mathrm{M}}=\left\vert 
\mathbf{P}_{1}\mathbf{P}_{2}\right\vert _{\mathrm{M}}=s  \label{a4.10}
\end{equation}%
is the length of vectors in the space-time of Minkowski, whereas their true
length in the space-time (\ref{a4.5}), (\ref{a4.6}) is%
\begin{equation}
\left\vert \mathbf{P}_{0}\mathbf{P}_{1}\right\vert =\left\vert \mathbf{P}_{1}%
\mathbf{P}_{2}\right\vert =\sqrt{\left\vert \mathbf{P}_{0}\mathbf{P}%
_{1}\right\vert _{\mathrm{M}}^{2}+2d\left( \sigma _{\mathrm{M}}\left(
P_{0},P_{1}\right) \right) }=\sqrt{s^{2}+2d\left( \sigma _{\mathrm{M}}\left(
P_{0},P_{1}\right) \right) }  \label{a4.11}
\end{equation}

According to (\ref{a1.1}), (\ref{a4.5}), we have%
\begin{equation}
\left( \mathbf{P}_{0}\mathbf{P}_{1}.\mathbf{Q}_{0}\mathbf{Q}_{1}\right)
=\left( \mathbf{P}_{0}\mathbf{P}_{1}.\mathbf{Q}_{0}\mathbf{Q}_{1}\right) _{%
\mathrm{M}}+w\left( P_{0},P_{1},Q_{0},Q_{1}\right)  \label{a4.12}
\end{equation}%
where $\left( \mathbf{P}_{0}\mathbf{P}_{1}.\mathbf{Q}_{0}\mathbf{Q}%
_{1}\right) _{\mathrm{M}}$ is the scalar product in the space-time of
Minkowski and%
\begin{equation}
w\left( P_{0},P_{1},Q_{0},Q_{1}\right) =d\left( \sigma _{\mathrm{M}}\left(
P_{0},Q_{1}\right) \right) +d\left( \sigma _{\mathrm{M}}\left(
P_{1},Q_{0}\right) \right) -d\left( \sigma _{\mathrm{M}}\left(
P_{0},Q_{0}\right) \right) -d\left( \sigma _{\mathrm{M}}\left(
P_{1},Q_{1}\right) \right)  \label{a4.14}
\end{equation}%
Thus, using relations (\ref{a4.11}),(\ref{a4.12}), one may write the
equivalence relations $\mathbf{P}_{0}\mathbf{P}_{1}$eqv$\mathbf{P}_{1}%
\mathbf{P}_{2}$ (\ref{a1.5}), (\ref{a1.6}) in terms of the Minkowskian world
function $\sigma _{\mathrm{M}}$ and distortion $d$.%
\begin{equation}
s\left( s+\alpha _{0}\right) +w\left( P_{0},P_{1},P_{1},P_{2}\right) =s^{2}
\label{a4.15}
\end{equation}%
\begin{equation}
\left( s+\alpha _{0}\right) ^{2}-\mathbf{\alpha }^{2}=s^{2},\qquad \mathbf{%
\alpha }^{2}=\alpha _{1}^{2}+\alpha _{2}^{2}+\alpha _{3}^{2}\geq 0
\label{a4.16}
\end{equation}%
where%
\begin{eqnarray}
w\left( P_{0},P_{1},P_{1},P_{2}\right) &=&d\left( \sigma _{\mathrm{M}}\left(
P_{0},P_{2}\right) \right) -d\left( \sigma _{\mathrm{M}}\left(
P_{0},P_{1}\right) \right) -d\left( \sigma _{\mathrm{M}}\left(
P_{1},P_{2}\right) \right)  \notag \\
&=&\lambda _{0}^{2}f\left( \frac{\left( 2s+\alpha _{0}\right) ^{2}-\mathbf{%
\alpha }^{2}}{2\sigma _{0}}\right) -2\lambda _{0}^{2}f\left( \frac{s^{2}}{%
2\sigma _{0}}\right)  \label{a4.17}
\end{eqnarray}%
Here four quantities $\alpha =\left\{ \alpha _{0},\mathbf{\alpha }\right\}
=\left\{ \alpha _{0},\alpha _{1},\alpha _{2},\alpha _{3}\right\} $ are to be
determined as a solution of the system of two equations (\ref{a4.15}), (\ref%
{a4.16}).

Equations (\ref{a4.16}) and (\ref{a4.15}) are written in the form 
\begin{equation}
2s\alpha _{0}+\alpha _{0}^{2}-\mathbf{\alpha }^{2}=0  \label{a4.18}
\end{equation}%
\begin{equation}
s\alpha _{0}+\lambda _{0}^{2}f\left( \frac{\left( 2s+\alpha _{0}\right) ^{2}-%
\mathbf{\alpha }^{2}}{2\sigma _{0}}\right) -2\lambda _{0}^{2}f\left( \frac{%
s^{2}}{2\sigma _{0}}\right) =0  \label{a4.19}
\end{equation}

Resolving equation (\ref{a4.18}) in the form%
\begin{equation}
\alpha _{0}=-s+\sqrt{s^{2}+\mathbf{\alpha }^{2}}=\frac{\mathbf{\alpha }^{2}}{%
s+\sqrt{s^{2}+\mathbf{\alpha }^{2}}}  \label{a4.20}
\end{equation}%
and eliminating $\alpha _{0}$ from (\ref{a4.19}), we obtain%
\begin{equation}
\frac{s\mathbf{\alpha }^{2}}{s+\sqrt{s^{2}+\mathbf{\alpha }^{2}}}+\lambda
_{0}^{2}f\left( \frac{s\left( s+\sqrt{s^{2}+\mathbf{\alpha }^{2}}\right) }{%
\sigma _{0}}\right) -2\lambda _{0}^{2}f\left( \frac{s^{2}}{2\sigma _{0}}%
\right) =0  \label{a4.21}
\end{equation}%
After transformations we obtain%
\begin{equation}
\mathbf{\alpha }^{2}=\frac{\lambda _{0}^{2}}{s}\left( s+\sqrt{s^{2}+\mathbf{%
\alpha }^{2}}\right) \left( 2f\left( \frac{s^{2}}{2\sigma _{0}}\right)
-f\left( \frac{s\left( s+\sqrt{s^{2}+\mathbf{\alpha }^{2}}\right) }{\sigma
_{0}}\right) \right)   \label{a4.22}
\end{equation}%
Introducing dimensionless quantities $x,$ $l$ , $k$%
\begin{equation}
x=\frac{\mathbf{\alpha }^{2}}{\sigma _{0}},\qquad l=\frac{s}{\sqrt{\sigma
_{0}}},\qquad k=\frac{\lambda _{0}^{2}}{\sigma _{0}}  \label{a4.24}
\end{equation}%
we obtain 
\begin{equation}
x=k\frac{l+\sqrt{l^{2}+x}}{l}\left( 2f\left( \frac{l^{2}}{2}\right) -f\left(
l\left( l+\sqrt{l^{2}+x}\right) \right) \right)   \label{a4.25}
\end{equation}%
We are interested only in solutions $x=\frac{\mathbf{\alpha }^{2}}{\sigma
_{0}}\geq 0$. These solutions are possible only for some values of $l^{2}$.
Boundaries of the regions with the positive solutions of equation (\ref%
{a4.25}) are determined as zeros of the equation 
\begin{equation}
k\left( 2f\left( \frac{l^{2}}{2}\right) -f\left( 2l^{2}\right) \right) =0
\label{a4.26}
\end{equation}%
If the function $f$ is defined by the relation (\ref{a4.6})%
\begin{equation}
f\left( l^{2}\right) =\left\{ 
\begin{array}{lll}
1 & \text{if} & l^{2}>1 \\ 
l^{2} & \text{if} & l^{2}\leq 1%
\end{array}%
\right. ,  \label{a4.27}
\end{equation}%
zeros of equation (\ref{a4.26}) are $\left\{ 0,1\right\} $. It means that
values $l^{2}\in \left( 0,1\right) $ are prohibited. The values 
\begin{equation}
l^{2}\in \left( 0,1\right)   \label{a4.27a}
\end{equation}%
are prohibited in the sense, that evolution of a particle with such a value
of $l^{2}$ is impossible.

According to (\ref{a4.11}) and (\ref{a4.24}) the geometrical mass $\mu $ is
expressed as follows 
\begin{equation}
\mu =\left\vert \mathbf{P}_{0}\mathbf{P}_{1}\right\vert =\sqrt{\left\vert 
\mathbf{P}_{0}\mathbf{P}_{1}\right\vert _{\mathrm{M}}^{2}+2\lambda
_{0}^{2}f\left( \frac{l^{2}}{2}\right) }=\sqrt{\sigma _{0}l^{2}+2\lambda
_{0}^{2}f\left( \frac{l^{2}}{2}\right) }  \label{a4.28}
\end{equation}%
Then it follows from (\ref{a4.28}), that the following values of the
geometrical mass $\mu $%
\begin{equation}
\mu ^{2}\in \left( 0,\sigma _{0}\left( 1+k\right) \right) =\left( 0,\sigma
_{0}+\lambda _{0}^{2}\right)   \label{a4.29}
\end{equation}%
are prohibited, because for these values of $\mu ^{2}$ the quantity $\mathbf{%
\alpha }^{2}<0$. As one can see, the case $\sigma _{0}\rightarrow 0$ with
fixed $\lambda _{0}>0$ corresponds to the discrete space-time geometry (\ref%
{a4.0}) with the minimal size $\lambda _{0}$. In this case the geometrical
mass $\mu <\lambda _{0}$ is prohibited. Particles of the mass $\mu <\lambda
_{0}$ are impossible in the discrete space-time geometry (\ref{a4.0}), and
the geometric mass $\mu =\lambda _{0}$ is the minimal mass in the space-time
geometry (\ref{a4.0}). Thus, the case $k=\lambda _{0}^{2}/\sigma _{0}=\infty 
$ corresponds to the completely discrete geometry. It is reasonable to
consider the case of finite $k=\lambda _{0}^{2}/\sigma _{0}$ as the case of
partly discrete geometry. In this case the quantity $k=\lambda
_{0}^{2}/\sigma _{0}$ may be considered as a degree of discreteness of the
space-time geometry.

In general, the concept of partly discrete geometry is not conventional. But
it is absent in the conventional approach, because the conception of a
discrete geometry is not developed properly. However, it appears that one
can introduce such a continuous parameter $k$, that at $k=0$ we have a
continuous space-time geometry (geometry of Minkowski), and at $k=\infty $
we have a completely discrete geometry. Investigating the space-time
geometry with arbitrary $k$, we use general methods of T-geometry
investigation. In such a situation it seems to be reasonable to introduce
the concept of partly discrete geometry with the discreteness scale $\lambda
_{0}$ and the discreteness degree $k$ .

In the partly discrete multivariant geometry there are constraints on
possible mass of a particle even in the case, when the discreteness is not
complete. The discreteness of the space-time geometry is connected with the
capacity of discrimination, generated by the fact of zero-variance, when the
equations (\ref{a1.5}), (\ref{a1.6}), describing equivalence of two vectors,
have no solution. Zero-variance and multivariance are two sides of one coin.
Multivariance generates quantum effects, whereas the zero-variance generates
discrete character of properties of elementary particles. Quantum effects
are generated by large values of the world function ($\sigma >\sigma _{0}$),
whereas zero-variance and discreteness are generated by small values of the
world function ($\sigma <\sigma _{0}$).

Substituting (\ref{a4.20}) in (\ref{a4.8}), we obtain%
\begin{equation}
P_{0}=\left\{ 0,0,0,0\right\} ,\qquad P_{1}=\left\{ s,0,0,0\right\} ,\qquad
P_{2}=\left\{ s+\sqrt{s^{2}+\mathbf{\alpha }^{2}},\alpha _{1},\alpha
_{2},\alpha _{3}\right\}   \label{a4.30}
\end{equation}%
\begin{eqnarray}
\mathbf{P}_{0}\mathbf{P}_{1} &=&\left\{ s,0,0,0\right\} ,\qquad \mathbf{P}%
_{1}\mathbf{P}_{2}=\left\{ \sqrt{s^{2}+\mathbf{\alpha }^{2}},\alpha
_{1},\alpha _{2},\alpha _{3}\right\}   \label{a4.31} \\
\mathbf{P}_{0}\mathbf{P}_{2} &=&\left\{ s+\sqrt{s^{2}+\mathbf{\alpha }^{2}}%
,\alpha _{1},\alpha _{2},\alpha _{3}\right\}   \label{a4.32}
\end{eqnarray}

\section{Concluding remarks}

Consideration of T-geometry as a space-time geometry admits one to obtain
dynamics of a particle as corollary of the geometrical object structure.
Evolution of the geometrical object in the space-time is determined by the
skeleton $\left\{ P_{0},P_{1},..P_{n}\right\} $ of the geometrical object
and by fixing of the leading vector $\mathbf{P}_{0}\mathbf{P}_{1}$. The
skeleton and the leading vector determine the world chain, which describes
the evolution completely. The world chain may be wobbling, it is a
manifestation of the space-time geometry multivariance. Quantum effects are
only one of manifestations of the multivariance. It is remarkable, that for
determination of the world chain one does not need differential equations,
which may be used only in the space-time manifold. One does not need
space-time continuity (continual geometry). Of course, one may introduce the
continual coordinate system and write dynamic differential equation there.
One may, but it is not necessary. In general, geometrical dynamics (i.e.
dynamics generated by the space-time geometry) is a discrete dynamics, where
step of evolution is determined by the length of the leading vector. It is
possible, that one will need a development of special mathematical technique
for an effective use of the geometrical dynamics.

The real space-time geometry contains the quantum constant $\hbar $ as a
parameter. As a result the geometric dynamics explains freely quantum
effects, but not only them. The particle mass is geometrized (the particle
mass is simply a length of some vector). As a result the problem of mass of
elementary particles is simply a geometrical problem. It is a problem of the
structure of elementary geometrical object and its evolution. One needs
simply to investigate different forms of skeletons of simplest geometrical
objects. Besides, the space-time geometry, looking as a continuous geometry,
appears to be partly discrete. These discreteness of "continual" space-time
geometry is a source of a discrete character of the elementary particle
properties. In turn T-geometry (new conception of geometry) is a source of
multivariance and of the discrimination capacity (zero-variance), which are
responsible for quantum effects and the discrete character of the elementary
particles properties.

\end{document}